\documentclass[twocolumn,prb,showpacs,preprintnumbers,amsmath,amssymb]{revtex4}
\usepackage[dvipdfm]{graphicx}
\usepackage{dcolumn}
\usepackage{bm}
\usepackage{amssymb}

\begin{document}
\title{Phase diagrams of $XXZ$ model on a depleted square lattice}

\author{Kwai-Kong Ng}
\affiliation{Department of Physics, Tunghai University, Taichung, Taiwan}
\date{\today}
\begin{abstract}
Using quantum Monte Carlo (QMC) simulations and a mean field theory, we investigate the spin-1/2 $XXZ$ model with nearest-neighbor interactions on a periodic depleted square lattice. In particular, we present results for 1/4 depleted lattice in an applied magnetic field and investigate the effect of depletion on the ground state. The ground-state phase diagram is found to include an antiferromagnetic (AF) phase of magnetization $m_{z}=\pm 1/6$ and an in-plane ferromagnetic (FM) phase with finite spin stiffness. The agreement between the QMC simulations and the mean field theory based on resonating trimers suggests the AF phase and in-plane FM phase can be interpreted as a Mott insulator and superfluid of trimer states, respectively. While the thermal transitions of the in-plane FM phase are well described by the Kosterlitz-Thouless transition, the quantum-phase transition from the AF phase to in-plane FM phase undergo a direct second-order insulator-superfluid transition upon increasing magnetic field. 
 \end{abstract}

\pacs{75.10.Jm, 75.45.+j, 05.30.Jp, 75.40.Mg}

\maketitle
\section{Introduction}
It is well known that depletion would weaken the long-range spin correlation in various quantum-spin systems. \cite{Dagotto,Zeng,Troyer,Liu} Reduced dimension arising from depletion would enhance quantum fluctuations in the system that consequently diminishes the spin-spin correlation in the long-range regime. It is known that the zero-point fluctuations in one-dimensional spin-1/2 Heisenberg model lead to power law decay in the spin-spin correlation and therefore destabilize the classic Neel state. Depletion of sites in two-dimensional magnetic systems would, therefore, ultimately destroy the long-range order (LRO) in the ground state by lowering dimensionality. One example is the destruction of the LRO of triangular spin-1/2 Heisenberg antiferromagnet by depleting 1/4 of the spins that leads to a Kagome lattice.\cite{Zeng} Similar phenomenon also occurs in triangular Bose-Hubbard model. The LRO phase in the triangular lattice \cite{Wessel1} appears when nearest-neighbor (nn) repulsive coupling $V \gtrsim 5.13$ but can only be stabilized for larger $V \gtrsim 8$ in the Kagome lattice \cite{Isakov,Damle} because of the enhanced quantum fluctuations caused by the site depletion. The LRO phase also emerges as an interesting valence-bond solid where bosons are delocalized around a subset of hexagons in the depleted system. Unlike these frustrated systems, another experimental important example is the 1/5 depleted Heisenberg model of CaV$_4$O$_9$ where the increased quantum fluctuations are not sufficient to destroy the LRO. \cite{Taniguchi,Ueda,Troyer} Heisenberg model of 1/4 depleted square lattice has also been investigated. \cite{Liu} Although spin correlations are weakened due to the depletion, quantum fluctuations, again, are not strong enough to destabilize the LRO. Given its simple lattice structure, it is surprising to note that a complete investigation of the  $XXZ$ model in the 1/4 depleted square lattice is still missing. 

On the other hand, intense efforts have recently been focused on the exotic supersolid \cite{Penrose,Andreev} phase on a variety of models. \cite{Wessel1,Batrouni,Schmid,Sengupta,Ng,Laflorencie}. One natural question to ask is whether depletion would, via reducing diagonal LRO, allow the possibility of coexistence of the off-diagonal and diagonal LROs. Therefore, it will be worthy to address this issue in the simplest depleted system. It is the object of this work to investigate the role of depletion on the ground-state phase diagram of the 1/4 depleted $XXZ$ model by employing both the quantum Monte Carlo (QMC) simulation and the timer basis mean field (MF) theory.

\section{Model}
\begin{figure}
\includegraphics[width=4cm]{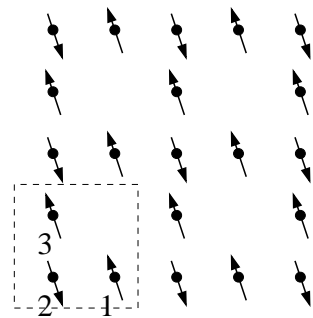}
\caption{The unit cell and site indices of the 1/4 depleted square lattice. The spin pattern represents the $m_z=1/6$ AF phase.}
\label{lattice}
\end{figure}

The lattice structure and the unit cell of the system are shown in Fig.\ref{lattice}. Each unit cell contains three spins with two of which (sites 1 and 3) are equivalent by symmetry. As shown in the figure, site 2 connects to sites 1 and 3 and has the coordination number $z=4$, while for the sites 1 and 3 $z=2$. This difference in coordination number $z$ leads to distinct chemical potentials $\mu$ on these sites when transformed the $XXZ$ spin model into the hardcore Bose-Hubbard model.

In this study, we consider the standard spin-1/2 $XXZ$ Hamiltonian:
\begin{equation}
H_{XXZ} =  \sum_{i,j}^{nn} \left( \Delta S^z_i S^z_j + \frac{J}{2} ( S^+_i S^-_j + h.c.) \right) - h\sum_iS^z_i
\end{equation}

\noindent in the depleted square lattice. Here the common notations of spin operators $S^z, S^\pm$ are used. To facilitate further analysis, $H_{XXZ}$ can be decomposed into the inter-cell and intra-cell parts and is written as:

\begin{eqnarray}\label{Hs}
H_s  &=&  \sum_{i} \left( H_{ai} + H_{bi} \right)\\
H_{ai} &=& \Delta S^z_{i2} (S^z_{i1} + S^z_{i3}) + \frac{J}{2} S^+_{i2} (S^-_{i1} + S^-_{i3}) + h.c. \nonumber \\ 
&& - h \sum_{\alpha} S^z_{i\alpha}  \nonumber \\
H_{bi} &=&  \Delta S^z_{i2} (S^z_{i+\hat{x} 1} + S^z_{i+\hat{y} 3})  + \frac{J}{2} S^+_{i2} (S^-_{i+\hat{x} 1} + S^-_{i+\hat{y} 3})  \nonumber \\
&& + h.c. \nonumber
\end{eqnarray}

\noindent where $H_{ai}$ and $H_{bi}$ are the Hamiltonians responsible for the inter-cell and intra-cell couplings, respectively.  $\sum_{i}$ now sums over all unit cells and $\sum_{\alpha}$ sums over the sites within a cell. $H_s$ is equivalent to a hardcore boson Hubbard model under the transformations 
$ S^+_{i\alpha}=b^+_{i\alpha} $ and  $S^z_{i\alpha}=n_{i\alpha}-\frac{1}{2} $. such that:

\begin{eqnarray}
H_B  &=&  \sum_{i} \left( H'_{ai} + H'_{bi} + \frac{1}{2}(\Delta+3h)\right)\\
H'_{ai} &=& V n_{i2} (n_{i1} + n_{i3}) + \frac{t}{2} b^+_{i2} (b^-_{i1} + b^-_{i3}) + h.c. \nonumber \\
&& - \sum_{\alpha}\mu_{\alpha} n_{i\alpha} \nonumber \\
H'_{bi} &=&  V n_{i2} (n_{i+\hat{x} 1} + n_{i+\hat{y} 3})  + \frac{t}{2} b^+_{i2} (b^-_{i+\hat{x} 1} + b^-_{i+\hat{y} 3}) \nonumber \\
&& + h.c., \nonumber
\end{eqnarray}

\noindent in which $\mu_{1}=\mu_{3}=h+\Delta$ and $\mu_{2}=h+2\Delta$ due to the distinct coordination number $z$ for the two types of sites. Hereafter, we will mainly focus on the discussion of the spin model but will also use the boson language if appropriate.

\section{Quantum Monte Carlo}
The spin $XXZ$ Hamiltonian can efficiently be simulated by the standard stochastic series expansion (SSE)  (Ref. 17) approach without difficulties. Most of the results are done on lattices of 24x24 unit cells and up to 36x36 unit cells are considered in the study of quantum-phase transitions. Temperatures are chosen to be inversely proportional to the lattice size such that $T\varpropto1/(2L)$. The essential order parameters include the spin stiffness $\rho_s$ (or superfluidity in boson language) that signals the in-plane FM phase. It is related to the winding number fluctuation of the updated loops in the SSE algorithm and can be easily implemented. \cite{Pollock} The broken symmetry of the lattice structure is reflected by the measured spin-structure factor 

\begin{equation}
 S(\textbf{Q})/N=\frac{1}{N^2}\sum_{ij}\langle S^z_i S^z_j e^{i \textbf{Q}\textbf{r}_{ij}}\rangle
\end{equation}
in the simulation. We set the lattice constant of each unit cell as unity and focus only on the wave vectors $\textbf{Q}_0=(2\pi,2\pi), (2\pi,0)$ and $(0,2\pi)$ by which the lattice symmetry of the depleted lattice is characterized.

\section{Ground-state phase diagram}

\begin{figure}
\includegraphics[width=7cm]{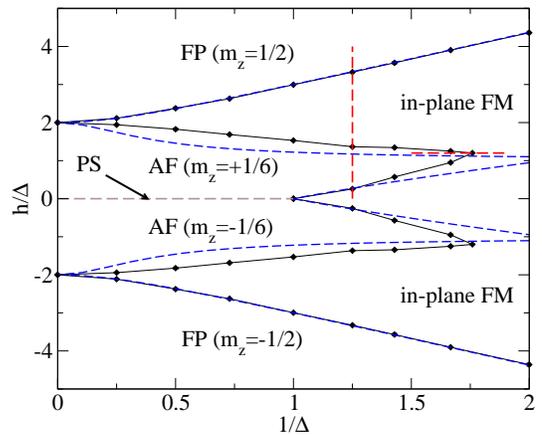}
\caption{(Color online) The ground-state phase diagram obtained by the QMC (denoted by black dots) simulations of lattice size 28x28. AF phase, characterized by the magnetization plateau $m_z=\pm1/6$, extends further into the in-plane FM phase as explained in the text. Results of a trimer-basis MF theory (blue dashed line) agree very well to the QMC data. For $\Delta<1$, phase separation (PS) is observed at $h=0$ while the ground state is a uniform superfluid for $\Delta>1$. FP denotes the fully polarized states for large fields. The vertical red line shows the scan for the order parameters in the Fig.\ref{param}.}
\label{phase}
\end{figure}

\begin{figure}
\includegraphics[width=7cm]{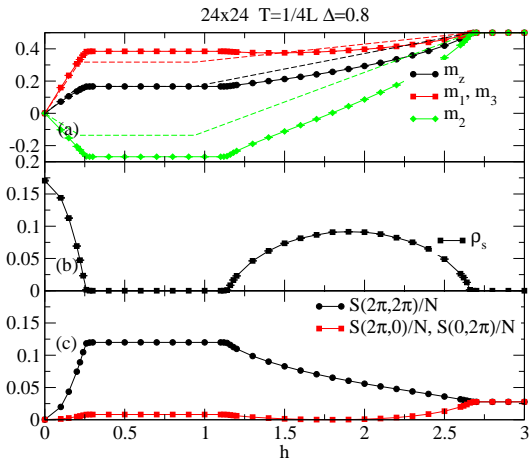}
\caption{(Color online) QMC data of order parameters of the (a) total and sublattice magnetizations, (b) spin stiffness $\rho_s$, and (c) structure factors as a function of field $h$ for $\Delta=0.8$. The dashed lines in (a) denote the corresponding mean-field magnetizations.}
\label{param}
\end{figure}

The ground-state phase diagram of hardcore Bose-Hubbard model in the undepleted square lattice \cite{Batrouni} contains simply a lobe of half-filled solid of checkerboard structure ($m_z=0$ Neel state) surrounded by a superfluid phase. Upon depletion of 1/4 sites, the $m_z=0$ Neel phase splits into two antiferromagnetic (AF) phases with magnetization $m_z=\pm1/6$ due to one missing spin in each unit cell as illustrated in the Fig. \ref{lattice}. As a consequence, the original $m_z=0$ lobe in the ground-state phase diagram now splits into two AF lobes in the depletion system as shown in the Fig. \ref{phase}. Due to the geometrical frustration at this magnetization, there is no 
$m_z=0$ state at zero field, and instead, a phase separation is observed for $\Delta>1$. For smaller $\Delta$, however, the zero-field ground state is a uniform in-plane FM characterized by a finite spin stiffness $\rho_s$. In contrast to the undepleted lattice, in which the AF LRO occurs only for $\Delta \eqslantgtr 1$, the AF phase in Fig. \ref{phase} extends further into the $\Delta<1$ regime and leads to the re-entrance of the in-plane FM phase when scanning along the increasing $h$ (see Fig. \ref{param}). This remarkable result indicates the regular depletion in our model in effect surpasses the quantum fluctuations and enhanced the AF LRO for smaller $\Delta \gtrsim 0.57$. Furthermore, the quantum-phase transition from AF to in-plane FM is continuous, in great contrast to the first-order phase transition from Neel to FM phases appeared in the undepleted square lattice. We stress that these observations are different from the case of the related Kagome lattice, which is essentially a regularly depleted triangular lattice. The region of the solid phase shrinks in that case and the solid to superfluid phase transition remains first order. Detail discussion of phase transitions will be offered in Sec. \ref{QPT}. In Fig.\ref{param} we present the order parameters as a function of $h$ with a typical $\Delta=0.8$.

The average magnetization $m_z$ increases monotonically as $h$ except within the AF phase which exhibits as a plateau of fixed $m_z=1/6$ as plotted in the Fig. \ref{param} (a). To gain further insight, the sublattice magnetizations $m_\alpha$ is also measured. The plateau phase has a finite-structure factor at $\textbf{Q}_0=(2\pi,2\pi)$, which represents the checkbroad crystal structure, while the spin stiffness $\rho_s$ vanishes. Sublattice magnetizations $m_1$ and $m_3$ are identical but are different from the $m_2$ due to the distinct geometry. 
Away from the AF plateau phase, $\rho_s$ are finite for all $h$ smaller than the critical field $h_{c3}$, beyond which all spins are fully polarized. However, these in-plane FM (superfluid) states are not uniform in space as illustrated by the finite structure factors $S(\textbf{Q}_0)$ and the distinct sublattice magnetizations. The superfluid condensates vary at site 2 and sites 1 and 3 such that the ground state is a non-uniform superfluid. Nevertheless, one has to note that the finite values of $S(\textbf{Q}_0)$ only reflect the depletion-induced internal structure of the unit cells. The superfluid indeed does not break the crystal symmetry.  This differs from the usual spin supersolid where the crystal symmetry is broken in addition to the broken gauge symmetry. Therefore the regular depletion in our case has not lead to any supersolid phase in the ground-state phase diagram.

At zero field, all sublattice magnetizations are the same and it indicates a uniform FM or superfluid phase. To shed light on the ground-state wave function, we construct a simple trimer-basis MF theory which surprisingly, agrees very well to the numerical results.

\section{Mean-field theory}
We begin with the exact ground-state wave function of the spin trimers in a unit cell, i.e., ignoring the inter-cell coupling $H_{bi}$ in the original Hamiltonian $H_s$ in the Eq. \ref{Hs}. Denoting the spin-trimer basis as $|s_1 s_2 s_3\rangle$ with $s_i$ represents the spin at the site $i$, the four lowest-energy eigenstates of the trimers are: 

\begin{eqnarray*}
 \varphi_{+\frac{1}{2}} &=&  \frac{1}{\sqrt{1+\alpha^2}}  \left( | \uparrow \uparrow \downarrow \rangle  + | \downarrow \uparrow \uparrow  \rangle + \alpha|\uparrow \downarrow \uparrow \rangle  \right) \\
\varphi_{+\frac{3}{2}} &=& |\uparrow \uparrow \uparrow \rangle \\
\alpha&=&-\frac{1}{2} \left(\Delta + \sqrt{\Delta^2+8} \right)
\end{eqnarray*}

\noindent with eigenvalues
 \begin{eqnarray*}
\varepsilon_{+\frac{1}{2}}&=&\frac{\alpha}{2} - \frac{h}{2} \\
\varepsilon_{+\frac{3}{2}}&=&\frac{\Delta}{2} - \frac{3h}{2}
 \end{eqnarray*}

\noindent respectively, and the states $\varphi_{-\frac{1}{2}}$ and $\varphi_{-\frac{3}{2}}$ with spin up and down interchanged. For $0< h < (\Delta-\alpha)/2$, the resonating trimer states $\varphi_{+\frac{1}{2}}$ and $\varphi_{-\frac{1}{2}}$ are the lowest energy and the first excited state, respectively. Up to now the analysis is exact. These states are no longer eigenstates, however, once the inter-cell coupling $H_{bi}$ is taken into account. To proceed we neglect the higher-energy states and construct a superfluid wave function of the trimer states $\varphi_{\pm\frac{1}{2}}$ such that:

\begin{equation}
 \Psi_0 = \prod_i \left( u_0 \varphi_{+\frac{1}{2}i} + v_0 e^{i\theta} \varphi_{-\frac{1}{2}i} \right)
\end{equation}

\noindent
for small external field $h$. Here $u_0$ and $v_0$ are related by $u_0^2+v_0^2=1$ and are determined by the minimization of the total energy $E_0=\langle \Psi_0 | H_s | \Psi_0 \rangle$. The global phase $\theta$ will not alter the total energy and is set to zero for convenience. If $\varphi_{-\frac{1}{2}i}$ is considered as the vacuum state, the creation of a $\varphi_{+\frac{1}{2}i}$ state is equivalent to adding a spin 1 hardcore boson and $\Psi_0$ represents a superfluid of those hardcore bosons. At zero magnetic field $h=0$, the occupation of both $\varphi_{\pm\frac{1}{2}i}$ states are identical with $u_0=v_0=\frac{1}{\sqrt{2}}$ that leads to the zero total magnetization ($m_z=0$). Increasing $h$ will raise the number of  $\varphi_{+\frac{1}{2}i}$ states by flipping spins in the $\varphi_{-\frac{1}{2}i}$ states. The condensate density, $\varpropto u_0 v_0$, reduces as $u_0$ increases and  $v_0$ decreases from $1/\sqrt{2}$. Ultimately, $v_0$ drops to zero at a critical field $h_{c1}$ as $\Psi_0$ represents a Mott insulator of $\varphi_{+\frac{1}{2}i}$  with a spin gap that manifests as the magnetization plateau of the AF state in Fig.\ref{param}. Higher external fields will lower the energy of $\varphi_{+\frac{3}{2}i}$ and finally close up the spin gap at $h=h_{c2}$. The ground state is then represented by:

\begin{equation}
 \Psi_1 = \prod_i \left( u_1 \varphi_{+\frac{1}{2}i} + v_1 e^{i\phi} \varphi_{+\frac{3}{2}i} \right),
\end{equation}

\noindent
as another superfluid (SF) of hardcore bosons $\varphi_{+\frac{3}{2}i}$. The occupation of $\varphi_{+\frac{3}{2}i}$, in which spins are all polarized, increases as $h$, until the system is fully polarized at $h_{c3}=\Delta-\alpha$. The critical fields $h_{c1}$, $h_{c2}$, and $h_{c3}$ define the MF phase boundaries which are plotted in the ground-state phase diagram (Fig. \ref{phase}) for comparison with QMC results. The agreement is good enough to suggest that the simple trimer basis MF theory is capable of capturing the essential low energy physics of the depleted system. It is noted that mean field calculations on the square and triangular lattices \cite{Murthy,Otterlo} also yield similar good agreements to the exact numerical results. Nevertheless, significant derivations are observed in the vicinity of the tip of the AF phase. This is attributed to the proximity of the energy levels of the $\varphi_{-\frac{1}{2}}$,   $\varphi_{+\frac{1}{2}}$, and $\varphi_{+\frac{3}{2}i}$ so that all of them should be taken into account in the wave function in order to achieve a better agreement with the QMC results.

This trimer-basis MF theory predicts all the transitions are continuous and of the SF-insulator type, in contrast to the Kagome lattice in which discontinuous transitions are found. The QMC results of the quantum-phase transitions will be addressed in the next section.

\section{Thermal and quantum-phase transitions}\label{QPT}
\begin{figure}
\includegraphics[width=7cm]{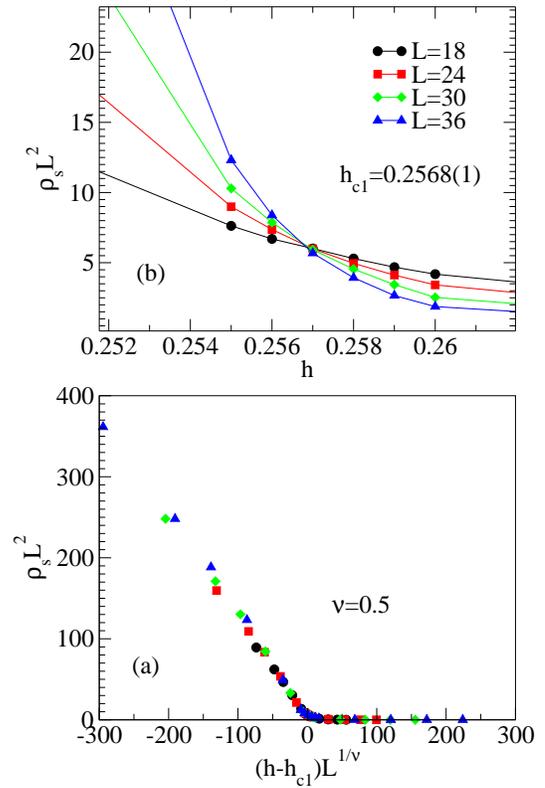}
\caption{(Color online) Finite-size scaling of $\rho_s$ at $h_{c1}$, the first critical field of in-plane FM to AF phase transition in Fig. \ref{param} for $\Delta=0.8$. (a) Data of $\rho_s L^2$ for different system sizes intersect at a critical field $h_{c1}=0.2568(1)$. (b) Data of $\rho_s L^2$ collapse well into a universal scaling curve when plotting against the $(h-h_{c1}) L^{1/\nu}$. The dynamical critical exponent is set to $z=0$ while correlation exponent $\nu=0$. The temperature used is $\beta=L^z/12$. The same for Fig. \ref{hc} and \ref{Dc}.}
\label{hc0}
\end{figure}

For the case of undepleted square lattice, \cite{Batrouni} the quantum-phase transitions from the in-plane FM phase to the AF phase (superfluid to solid phase in boson language) is apparently discontinuous because of the different broken symmetries in both phases. Similar situation occurs in the triangular lattice, \cite{Wessel1} albeit a supersolid phase exists in some small regimes of the phase diagram such that continuous quantum-phase transitions are observed from the supersolid phase to the superfluid phase. Away from the supersolid phase, direct transitions from the superfluid phase to solid phase are of first order. Upon depletion, since the superfluid and solid phases break distinct symmetries, a weakly first-order phase transition is recently observed \cite{Isakov,Damle}. These results are in contrast to our findings of the depleted model $H_s$.

\begin{figure}
\includegraphics[width=7cm]{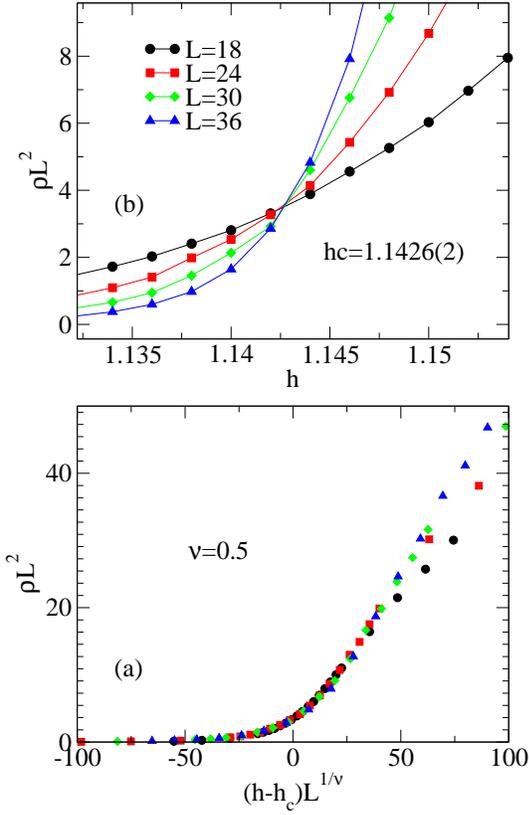}
\caption{(Color online) Finite-size scaling $\rho_s$ at $h_{c2}$, the second critical field of AF to in-plane FM phase transition in Fig. \ref{param} for $\Delta=0.8$. Data also appear to collapse for $z=2$ and $\nu=0.5$, though not as well as in Fig. \ref{hc0} for $h_{c1}$.}
\label{hc}
\end{figure}

To investigate the quantum-phase transitions in detail, we performed a finite size scaling (FSS) analysis for the continuous AF to in-plane FM phase transitions. In two-dimensions, close to a critical point $K_c$, $\rho_s$ scales as \cite{Fisher},

\begin{equation}\label{FS}
\rho_s= L^{-z}F_{\rho_s}[L^{1/\nu}(K-K_c),\beta/L^z]
\end{equation}

\noindent
where $L$ is the linear system size, $\nu$ the correlation exponent, $z$ the dynamical critical exponent, and $K_c-K$ is the distance to the critical point. Given the scaling relation, data of different system sizes $L$ should cross at the transition point $K_c$ when $\rho_s L^z$ is plotted against $K$ as long as $\beta$ changes as $L^{z}$ so that the second argument in $F_{\rho_s}$ remains constant. Equation \ref{FS} also implies the collapse of data onto a universal curve of the scaling function $F_{\rho_s}$ once $\rho_s L^z$ is plotted against $(K-K_c)L^{1/\nu}$ for the appropriate critical exponent $\nu$. For $\Delta=0.8$, we performed the finite-size scaling of $\rho_s$ at two critical magnetic fields $h_{c1}$ and $h_{c2}$. As illustrated in Figs. \ref{hc0}(a) and \ref{hc}(a), data of different system sizes intersect, respectively, at the same critical points of $h_{c1}=0.2568(1)$ and $h_{c2}=1.1426(2)$ for $z=2$. Furthermore, data of $\rho_s L^2$ in Figs. \ref{hc0}(b) and \ref{hc}(b) appear to collapse well for $\nu=0.5$ as expected from the superfluid-insulator universality class, \cite{Fisher} supports the notion of continuous phase transitions between AF and in-plane FM phases.

\begin{figure}
\includegraphics[width=7cm]{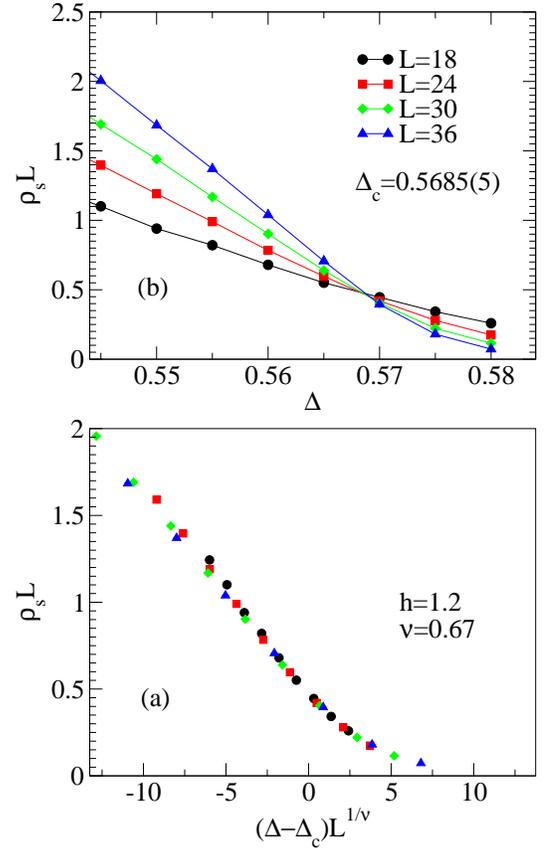}
\caption{(Color online) Finite-size scaling of $\rho_s$ crossing the tip of the lobe in Fig. \ref{phase} (red horizontal line in Fig. \ref{phase}) with $m_z=1/6$. Using $z=1$ and $\nu=0.67$, (a) the critical $\Delta_c$ is found to be $\Delta_c=0.5685(5)$ and (b) data collapse well onto the universal scaling curve.} 
\label{Dc}
\end{figure}

Similar analysis applies to the melting of AF phase to FM phase crosses the critical $\Delta_c$ (horizontal line in Fig. \ref{phase}). In Fig. \ref{Dc}, the QMC data scale well as stated in Eq. \ref{FS} for the critical exponents $z=1$ and $\nu=0.67$ with the critical $\Delta_c=0.5685(5)$. All these results suggest the observed quantum-phase transitions of AF to in-plane FM phases obey the Landau-Ginzburg-Wilson theory in 2D. In addition, the observation of a single peak in the histogram of QMC data at the critical points implies the possibility of weakly first-order phase transition is unlikely.

The critical behaviors of the depleted lattice can be understood as one noticed that there is in fact no broken-crystal symmetry in the AF phase. As mentioned in the Sec. IV, the finite-structure factors $S(\bf{Q_0})$ only reflect the internal spin structure within unit cells which originated from the two different types of sites in the lattice. Although $S(\bf{Q_0})$ are found finite in both phases, the crystal symmetry is indeed preserved. Therefore the phase transition is expected to be continuous between the insulating AF and the in-plane FM phase. On the other hand , the situation is different in the Kagome lattice, as well as the undepleted square lattice, where all sites are equivalent. The observed finite-structure factors indicate that the spin structure of the AF phase breaks the crystal symmetry of the system. Consequently, the phase transition to the in-plane FM phase is expected to be first order. This explains the distinct critical behaviors upon depletion in these closely related systems.

\begin{figure}
\includegraphics[width=7cm]{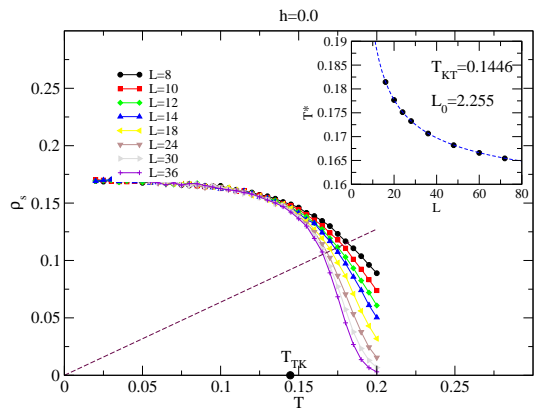}
\caption{(Color online) $\rho_s$ is plotted as a function of $T$ for different system sizes $L$ near the KT transition temperature $T_{KT}$ for $\Delta=0.8$ at $h=0$. The dashed line $\rho_s(T)=\frac{2}{\pi}T$ intersects $\rho_s$ curves of different $L$ at temperatures $T^*(L)$, which follows the logarithmic behavior given by Eq. \ref{KT} as shown in the inset. }
\label{Tc_h0}
\end{figure}

\begin{figure}
\includegraphics[width=7cm]{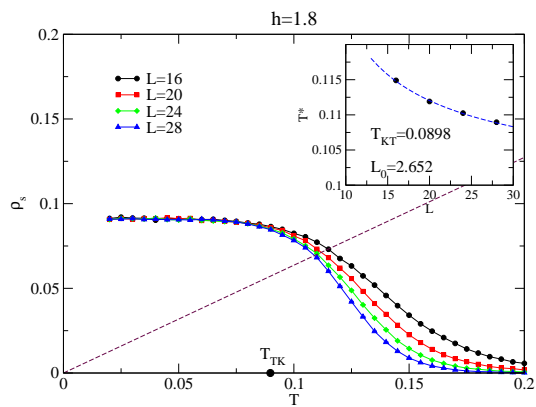}
\caption{(Color online) Same as Fig. \ref{Tc_h0} but now $h=1.8$.}
\label{Tc_h1}
\end{figure}

Besides the quantum-phase transitions, we also investigate the thermal phase transitions which are found to be well described by the Kosterlitz-Thouless (KT) theory \cite{Kosterlitz} in 2D. 
For $T>0$, only quasi-long-range order is possible in the $xy$ plane as continuous gauge symmetry cannot be broken in 2D. Nevertheless the spin stiffness $\rho_s$ can remain finite until reaching the KT temperature $T_{\texttt{KT}}$, which can numerically be determined by noting that the universal jump of $\rho_s$ at $T_{\texttt{KT}}$ is given by

\begin{equation}
\rho_s(T_{\texttt{KT}})=\frac{2}{\pi}T_{\texttt{KT}}.
\end{equation}

Taking into account the logarithmic finite size corrections, the intersections of the line $\rho_s(T)=\frac{2}{\pi}T$ and the QMC data for different system sizes occur at 

\begin{equation}\label{KT}
T^*(L)=T_{KT}\left(1+\frac{1}{2 \ln(L/L_0)}\right).
\end{equation} 

\noindent
As shown in Figs. \ref{Tc_h0} and \ref{Tc_h1} for $h=0$ and $1.8$, respectively for $L$ up to 36, the intersections $T^*$ follows well the logarithmic correction of Eq. \ref{KT}. For infinite system, $\rho_s$ will vanish at the measured $T_{\texttt{KT}}$ and the system becomes paramagnetic above $T_{\texttt{KT}}$. 

\section{Summary}
A comprehensive study in the ground-state phase diagram of the 1/4 depleted square lattice is given in this work. The phase diagram contains an in-plane FM phase and a AF phase, where the AF phase extends further than the case in the undepleted lattice due to the regularity of the depletion. In the in-plane FM phase, in spite of the coexistence of finite-structure factor $S(\textbf{Q}_0)$ and $\rho_s$, the ground state does not break the crystal symmetry and is not a supersolid state. Quantum-phase transitions from the AF to in-plane FM are demonstrated to be continuous, in stark contrast to the case of undepleted lattice. The critical exponents $\nu$ and $z$ retrieved from the FSS are consistent with the prediction from the superfluid-insulator universality class. On the other hand, numerical data of the thermal phase transitions belong to the KT type as expected for the 2D superfluids. Apart from the QMC results, a trimer-basis MF theory is proposed to gain more insights of the model. Based on the trimer states of each isolated unit cells, the MF theory projects out higher-energy states and attempts to capture essential physics of the system by retaining the lower-energy states. The AF is represented as $\prod_i  \varphi_{+\frac{1}{2}i}$ and the FM can be expressed as superfluids of $\varphi_{+\frac{1}{2}}$ in the background of $\varphi_{-\frac{1}{2}}$ or  $\varphi_{+\frac{3}{2}}$. The agreement between QMC and MF results is remarkable, given the simplicity of the trimer-basis MF theory. 

While a triple laser beam design in cold atom gases \cite{Santos}  has been proposed to simulate a Kagome lattice, whether similar techniques may apply to the 1/4 depleted square lattice is worthy to pursue. Furthermore, this model also related to the Bose-Hubbard model with both nn and next-nearest-neighbor couplings. Recent study on this model found a commensurate SS of density 1/4, which has similar structure of the 1/4 depleted lattice studied here \cite{Ng1}. One interesting issue is if the nnn coupling (i.e. diagonal bonds in our case) would lead to unexpected types of quantum phases. Moreover, adding only half of the diagonal bonds in the 1/4 depleted lattice leads to the Kagome lattice in which very different ground states, including VBS solid, is observed. The investigation of how do the two distinct phase diagrams merge as the diagonal coupling is increased will also be important for the understanding of phase transitions of VBS solids. 

\begin{acknowledgments}
We thank M. F. Yang and Y. C. Chen for helpful discussions. The numerical computations are performed in the Center for High Performance Computing of the THU. 
 K.K.N. acknowledges financial support by the NSC
(R.O.C.), under Grant No. NSC 97-2112-M-029-003-MY3 and No. NSC 95-2112-M-029-010-MY2.
\end{acknowledgments}


\begin{thebibliography}{99}
\bibitem{Dagotto}
E. Dagotto and T. M. Rice, Science, {\bf 271}, 618 (1996).

\bibitem{Zeng}
C. Zeng and V. Elser, Phys. Rev. B {\bf 42}, 8436 (1990).

\bibitem{Troyer}
M. Troyer, H. Kontani and K. Ueda, Phys. Rev. Lett. {\bf 76}, 3822 (1996).

\bibitem{Liu}
Y. J. Liu, Y. C. Chen, M. F. Yang and C. D. Gong, J. Phys. {\bf 18} 1805, (2006).

\bibitem{Wessel1}
S. Wessel and M. Troyer, Phys. Rev. Lett. {\bf 95}, 127205 (2005);
D. Heidarian and K. Damle, ibid {\bf 95}, 127206 (2005); R. G. Melko, A. Paramekanti, A. A. Burkov, A. Bishwanath, D. N. Sheng, and L. Balents, ibid {\bf 95}, 127207 (2005).

\bibitem{Isakov}
S. V. Isakov, S. Wessel, R. G. Melko, K. Sengupta, and Y. B. Kim, Phys. Rev. Lett. {\bf 97}, 147202 (2006).

\bibitem{Damle}
K. Damle and T. Senthil, Phys. Rev. Lett. {\bf 97}, 067202 (2006).

\bibitem{Taniguchi}
S. Taniguchi, T. Nishikawa, Y. Yasui, Y. Kobayashi, M. Sato, T. Nishioka, M. Kontani, and K. Sano, J. Phys. Soc. Jpn. {\bf 64}, 2758 (1995).

\bibitem{Ueda}
K. Ueda, H. Kontani, M. Sigrist and P. A. Lee, Phys. Rev. Lett. {\bf 76}, 1932 (1996).

\bibitem{Penrose}
O. Penrose and L. Onsager, Phys. Rev. {\bf 104}, 576 (1956).

\bibitem{Andreev}
A. F. Andreev and I. M. Lifshitz, Sov. Phys. JETP {\bf 29}, 1107
(1969); G. V. Chester, Phys. Rev. A, {\bf 2}, 256 (1970); A. J.
Leggett, Phys. Rev. Lett. {\bf 25}, 1543 (1970).


\bibitem{Batrouni}
G.G. Batrouni and R.T. Scalettar, Phys. Rev. Lett. {\bf 84}, 1599
(2000); F. Hebert, $et$ $al.$. Phys. Rev. B {\bf 65}, 014513 (2001).

\bibitem{Schmid}
G. Schmid, S. Todo, M. Troyer, and A. Dorneich, Phys. Rev. Lett. {\bf 88}, 167208 (2002).

\bibitem{Sengupta}
P. Sengupta, L. P. Pryadko, F. Alet, M. Troyer and G. Schmid, Phys. Rev. Lett. {\bf 94}, 207202 (2005).

\bibitem{Ng}
K. K. Ng and T. K. Lee, Phys. Rev. Lett. {\bf 97}, 127204 (2006).

\bibitem{Laflorencie}
N. Laflorencie and F. Mila, Phys. Rev. Lett. {\bf 99}, 027202 (2007).

\bibitem{Sandvik}
A. W. Sandvik, Phys. Rev. B {\bf 59}, R14157 (1999); ibid {\bf 56},
11678 (1997); O. F. Sylju\r{a}sen and A. W. Sandvik, Phys. Rev. E {\bf
66}, 046701 (2002).

\bibitem{Pollock}
E. L. Pollock and D. M. Ceperley, Phys. Rev. B{\bf 36}, 8343 (1987).

\bibitem{Murthy}
G. Murthy, D. Arovas, and A. Auerbach, Phys. Rev. B {\bf 55}, 3104 (1997).

\bibitem{Otterlo}
A. van Otterlo, K. Wagenblast, R. Baltin, C. Bruder, R. Fazio, and G. Schon, Phys. Rev. B {\bf 52}, 16176 (1995)
 

\bibitem{Fisher} M. P. A. Fisher, P. B. Weichman, G. Grinstein, and D. S. Fisher, Phys. Rev. B {\bf 40}, 546 (1989).

\bibitem{Kosterlitz}
J. M. Kosterlitz and D. J. Thouless, J. Phys. C {\bf 6}, 1181 (1973).

\bibitem{Santos}
L. Santos, M. A. Baranov, J. I. Cirac, H.-U. Everts, H. Fehrmann, and M. Lewenstein, Phys. Rev. Lett. {\bf 93}, 030601 (2004); B. Damski, H. Fehrmann, H.-U. Everts, M. Baranov, L. Santos, and M. Lewenstein, Phys. Rev. A {\bf 72}, 053612 (2005).

\bibitem{Ng1} 
K. K. Ng and Y. C. Chen, Phys. Rev. B {\bf 77}, 052506 (2008). K. K. Ng, Y. C. Chen and Y. C. Tzeng, arXiv:09082478.







\end{thebibliography}
\end{document}